\documentclass{article}
\usepackage{ijcai21}

\usepackage{times}

\usepackage{soul}
\usepackage{url}
\usepackage[hidelinks]{hyperref}
\usepackage[utf8]{inputenc}
\usepackage[small]{caption}
\usepackage{graphicx}
\usepackage{amsmath}
\usepackage{booktabs}
\usepackage{color}
\usepackage{amsthm,amsmath,amssymb}
\usepackage{mathrsfs}
\usepackage{soul}
\usepackage{subfigure}
\usepackage{comment}
\usepackage{multirow}

\title{Open-Set Crowdsourcing using Multiple-Source Transfer Learning}
\author{
Guangyang Han$^1$\and
Guoxian Yu$^{2,3}$\footnote{Contact Author}\and Lei Liu$^{2,3}$\and Lizhen Cui$^{2,3}$\and
Carlotta Domeniconi$^{4}$,
Xiangliang Zhang$^5$
\\
\affiliations
$^1$College of Computer and Information Sciences, Southwest University, China\\
$^2$School of Software, Shandong University, China\\
$^3$Joint SDU-NTU Centre for Artificial Intelligence Research, Shandong University, Jinan, China\\
$^4$Department of Computer Science, George Mason University, USA\\
$^5$Department of Computer Science, King Abudullah University of Science and Technology, Saudi Arabia\\

\emails
hanguangyang@email.swu.edu.cn,
\{gxyu, l.liu, clz\}@sdu.edu.cn,
carlotta@cs.gmu.edu,
xiangliang.zhang@kaust.edu.sa
}

\begin{document}

\maketitle

\begin{abstract}
We raise and define a new crowdsourcing scenario, \emph{open set crowdsourcing}, where we only know the general theme of an unfamiliar crowdsourcing project, and we don't know its label space, that is, the set of possible labels. This is still a task annotating problem, but the unfamiliarity with the tasks and the label space hampers the modelling of the task and of workers, and also the truth inference. We propose an intuitive solution, OSCrowd. First, OSCrowd integrates crowd theme related datasets into a large source domain to facilitate partial transfer learning to approximate the label space inference of these tasks. Next, it assigns weights to each source domain based on category correlation. After this, it uses multiple-source open set transfer learning to model crowd tasks and assign possible annotations. The label space and annotations given by transfer learning will be used to guide and standardize crowd workers' annotations. We validate OSCrowd in an online scenario, and prove that OSCrowd solves the open set crowdsourcing problem, works better than related crowdsourcing solutions.
\end{abstract}

\section{Introduction}
Crowdsourcing is a new computation model that employs a large number of ordinary Internet workers to work together on a large project. In recent years, with the explosion of machine learning, especially deep neural networks, the demand for labeled data has been ever-increasing. As such, how to harness crowdsourcing to provide annotated data for machine learning has become a research hot-spot \cite{li2017crowdsourced}. A series of crowdsourcing platforms, such as \textit{Amazon Mechanical Turk} and \textit{CrowdFlower}, have emerged and serve as the workplaces for crowd workers. In the typical crowdsourcing practice, we usually assume the label space of the crowd tasks is known or fixed \cite{li2017crowdsourced}; in other words, we need to predefine the labels. In this way, the crowdsourcing task is modeled as a closed set label assignment problem. But in practice, this assumption is often violated. Suppose we have collected a set of animal-themed pictures, which include tigers, lions, horses, cats, rhinos, etc.. If we want to classify all samples in detail, due to the open set of the label space, it's hard to setup the crowd task as there are so many alternative labels for these images. This \emph{open set} problem also hampers quality, budget and latency control of crowdsourcing, as it's difficult and costly for crowd workers to go through all the options to choose the most appropriate label, and it's also difficult for requester to infer the task truth and the worker ability model when the annotation data is sparse. On the other hand, if we let the crowd workers give out labels instead of choosing from the given options, there may be a label inconsistency problem, which requires the design of additional tasks to further determine which labels are actually the same.

Transfer learning can transfer knowledge from related fields to help the learning model achieve a better performance in the target domain \cite{zhuang2020comprehensive}. Homogeneous transfer learning assumes that the source and target domains share the same feature and label space, and their difference lies in the marginal distribution of the feature space and/or the feature-label conditional distribution \cite{long2013transfer}. The recently studied open set transfer learning \cite{panareda2017open} and partial transfer learning \cite{zhang2018importance} relax the restriction of sharing the same label space for the source and target domains, thus making transfer learning more practical. Inspired by this, if we can transfer available knowledge from related domains to model target crowdsourcing tasks, and roughly infer the possible label space of target tasks, then we can better tackle an open set crowdsourcing project, and complete it with a lower budget and a lower degree of uncertainty.

To implement the above idea, we introduce a solution called Open-Set Crowdsourcing (\textbf{OSCrowd}). First, due to the lack of sufficient knowledge of target crowdsourcing tasks, OSCrowd collects as many relevant datasets as possible to ensure the coverage of the label space of target crowdsourcing tasks. Then, it integrates the datasets into a large source domain, models the problem as a partial transfer learning problem, infers the possible label space of target tasks, and weighs each source dataset. After this, it applies open set transfer learning to predict the labels of crowdsourcing tasks, combines weight information to give the machine annotating results of target tasks. Finally, OSCrowd uses the label space information and machine annotations to assist the design and execution of crowdsourcing workflow of target tasks.

The contributions of our work are summarized as follows:
{
\begin{itemize}
    \item We introduce and define the open set crowdsourcing annotating problem, discuss its relevance and challenges.
    \item We propose to use transfer learning to infer the label space of crowdsourcing tasks, and propose a solution to the open set crowdsourcing annotation problem.
    \item We simulate and verify the effectiveness of our solution in online crowdsourcing scenarios. OSCrowd solved the problem of open set crowdsourcing annotation problem, and achieved the effect of SOTA.
\end{itemize}
}

\section{Related Work}
Our work builds on crowdsourcing and transfer learning. A comprehensive coverage of these two areas is out of the scope of this paper. In the following, we introduce related work which is most relevant to our approach.
%Below are some background and related work involved in this paper.

\textbf{Crowdsourcing} is a new computation model that coordinates the crowd (Internet workers) to do micro-tasks in order to solve computer-hard problems; as an example, the world’s largest and most popular online encyclopedia, Wikipedia, is the product of many volunteers crowd workers. The task common to the different crowdsourcing scenarios is data processing \cite{li2017crowdsourced}, which includes collection, classification, modification, and reorganization, with data classification (labeling) being the most typical one.

The research on crowdsourcing mainly focuses on three aspects: quality, budget, and latency control \cite{li2017crowdsourced}. No worker is perfect, so the annotations provided by crowd workers inevitably contain noise. How to identify and reduce the impact of noise is a key quality control task \cite{zheng2017truth}. Crowd workers are generally not free, and the scale of crowdsourcing tasks is typically huge; in addition, repeated annotations are needed to ensure quality. As such, budget control is also very important, and this is usually managed by modeling tasks/workers to reduce the number of required annotations \cite{ho2013adaptive,li2016crowdsourcing,Tu2020CrowdWT}. For latency control, researchers reduce the total time required to complete crowdsourcing via improved task assignment or workflow design \cite{haas2015clamshell}.
%{\color{red}[how to make it? summarize the key techniques?]} \cite{haas2015clamshell,verroios2015tdp}.

However, most of the aforementioned worker/task modeling and truth inference methods are based on a closed label set. For example, they usually assume that the task is binary or an $n$-way classification problem. When the problem is open set, the quality of workers and difficulty of tasks are harder to estimate due to sparse annotated data and the label open set.

\textbf{Domain adaptation} is a common paradigm of homogeneous transfer learning. It studies the situation in which the feature and label spaces of the source and target domains are uniform, but the feature distribution (marginal distribution $P(X)$) is inconsistent \cite{blitzer2006domain}. Domain adaptation tries to transfer knowledge from the source domain to the target domain by aligning the feature spaces and by finding what they have in common \cite{ben2007analysis}. It is usually assumed that the source domain has rich supervised information, while the target domain has little or no label information. In recent years, deep learning, especially adversarial learning \cite{goodfellow2014generative}, has achieved excellent results in domain adaptation. Deep Domain Confusion (DDC) \cite{tzeng2014deep} adds a domain adaptation layer using the Maximum Mean Discrepancy (MMD) distance metric in AlexNet to help the network extract the common feature representation. Unsupervised Domain Adaptation by Backpropagation (UDAB) \cite{ganin2015unsupervised} sets up a domain classifier to force the network to learn domain-invariant feature representations. \cite{ganin2016domain,hoffman2018cycada} also adopt the idea of adversarial learning to bridge the differences between domains.

%{\color{red} [to do what? this sentence is not so complete and informative.] In addition, the connection with open set crowdsourcing and multi-source transfer are also not so clear. The related works should drive our research ideas and techniques.}.

\textbf{Partial \& Open-set transfer} relaxes the assumption that the label space is consistent, thereby making the transfer model more extensive and practical. Partial transfer assumes that the target domain label space is a subset of the source domain, while open-set transfer assumes the two label spaces partially overlap, and both have their own unique classes. \cite{zhang2018importance} adopted a domain adversarial network to weight source domain samples whose classes are likely to appear in the target domain. \cite{cao2019learning} further introduced two additional auxiliary discriminators to make the weights more reasonable. In open-set transfer, \cite{panareda2017open} assigns labels to target domain samples by calculating the sample-class center distance after the domain adaptation process is done.

% they both assume that the size and content of the target domain label space are unknown {\color{red}[original transfer can also handle unknown labels in the target domain, transfer can be from label, ]}.

%Multi-source transfer has multiple source domains.
In practice, it is difficult to find a source domain that completely covers the label space of the target domain, but it is easy to find multiple related source domains. Our idea is that, instead of using multiple homogeneous source domains for open set domain adaptation, we can combine multiple source domains as a whole, and then perform partial domain adaptation. In this way, we can unify multiple source domains into one framework to facilitate the transfer of knowledge.

\section{Methods}

\begin{figure*}[h!tbp]
\centerline{\includegraphics[width=18cm]{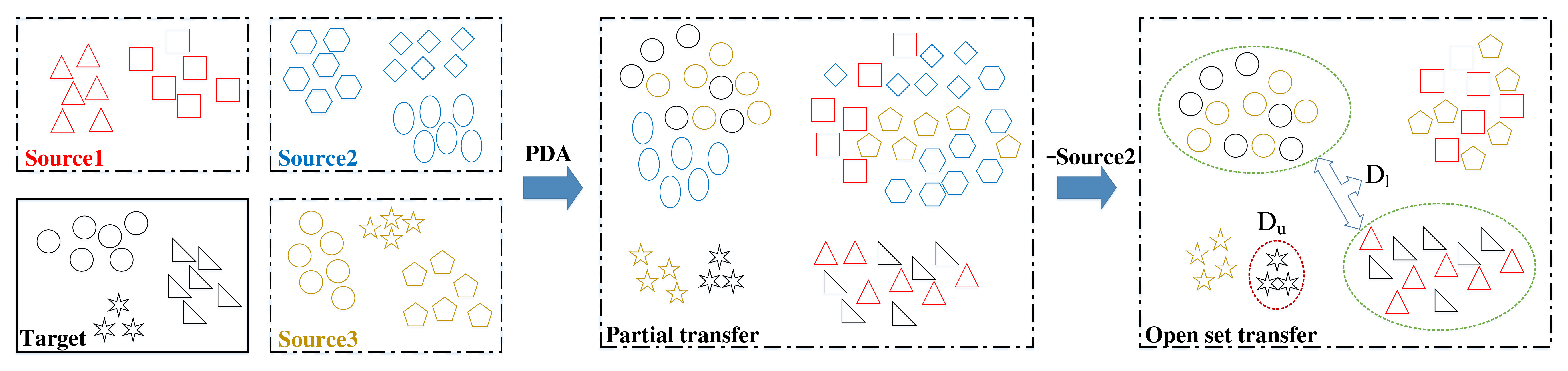}}
\vspace{-0.7em}
\caption{OScrowd first adopts partial domain adaptation (PDA) and GAN to align the feature space of the multiple-source domains and the target domain, considers the possibility of source domain classes being present in the target domain, and then excludes irrelevant source domains (Source2 in figure). Next, OScrowd uses open set transfer to assign target domain samples with one of the classes in the source domain, or to label them as unknown. The labeled samples correspond to simple tasks ($\mathcal{D}_l$), while the unknown tasks are considered more difficult ($\mathcal{D}_u$). $\mathcal{D}_l$ and $\mathcal{D}_u$ are then assigned to different quality workers for crowdsourcing. \textit{Best viewed in color.}}
\end{figure*}

\subsection{Notation and Problem Formulation}
In this section, the necessary definitions pertaining crowdsourcing and transfer learning are given \cite{zhuang2020comprehensive}.

\textit{Definition 1 (Task):} A task (in machine learning) $\mathcal{T}$ contains a label space $\mathcal{Y}$ and a decision function $f$ which needs to be learned from the training data, that is $\mathcal{T} = \{\mathcal{Y},f\}$. In crowdsourcing, the term \textit{task} usually refer to (a batch of) sample(s) assigned to crowd workers.

\textit{Definition 2 (Domain):} A domain consists of two parts, a feature space $\mathcal{X}$ and a marginal distribution $P(X)$.
%, the symbol $X$ denotes instance set, \textit{i.e.}, $X = \{\mathbf{x}|\mathbf{x}_i \in \mathcal{X}\}$
In practice, a domain is often represented by an instance set with or without labels. For example, a source domain is defined as $\mathcal{D}_s = \{ (\mathbf{x}_i,y_i)| \mathbf{x}_i \in \mathcal{X}^s, y_i \in \mathcal{Y}^s, i = 1, \cdots, n^s\}$.

\textit{Definition 3 (Transfer Learning):} Transfer learning uses knowledge implied in the source domains and source tasks ($\{\mathcal{D}_{s_i}, \mathcal{T}_{s_i}\}$) to improve the performance of the decision function $f^t$ on the target domain. Domain adaptation is a paradigm for transfer learning which focuses on aligning $P(X)$ for the source and the target domains.

\textit{Definition 4 (Open set Crowdsourcing):} We model Open Set Crowdsourcing as a target domain $\mathcal{D}_t = \{\mathbf{x}| \mathbf{x}_i \in \mathcal{X}^t, i=1, \cdots, n^t\}$, where $n^t$ is the number of samples to be annotated by a group of $W$ crowd workers $\mathcal{W} = \{w_i\}_{i=1}^{W}$ (the annotation is represented as $\mathbf{a}$). In open set crowdsourcing, we have some information about the label space $\mathcal{Y}^t$, but its size $n_y^t=|\mathcal{Y}^t|$ and the specific content are unknown.

Suppose we have a crowdsourcing project $\mathcal{D}_t$, of which we only know the general theme, such as the fact that it pertains a set of pictures about livestock. The first thing we need to do is to gather labeled datasets about livestock from the Internet, annotating them as $\{\mathcal{D}_{s_i}\}_{i=1}^{m'}$. Each dataset may share $(0 \sim n_y^t)$ classes with $\mathcal{D}_t$. We merge the datasets into a large source domain $\mathcal{D}_s$, and the label space of $\mathcal{D}_s$ should contain the label space of $\mathcal{D}_t$. We aim to: (i) discover the classes shared by $\mathcal{D}_s$ and $\mathcal{D}_t$; (ii) distinguish their differences; and (iii) transfer knowledge from $\mathcal{D}_s$ to help model the tasks in $\mathcal{D}_t$.
%\st{which can be achieved by \textit{Generative Adversarial Networks} (GAN)} \cite{goodfellow2014generative} and domain adaptation.

\subsection{Partial Domain Adaptation}
To meet the above three needs, inspired by the adversarial learning based domain adaption solutions \cite{ganin2016domain,tzeng2017adversarial}, OSCrowd adopts a domain adversarial network built on GAN \cite{goodfellow2014generative}. GAN originally contains two main parts, a generator ${G}$ and a discriminator ${D}$. Here, we consider three main parts for a domain adversarial network: two feature extractors and a domain discriminator. The feature extractors ${F}_s$ and ${F}_t$ for $\mathcal{D}_s$ and $\mathcal{D}_t$ can be the same or different; ${F}_s$ is trained on $\mathcal{D}_s$ and kept unchanged, while ${F}_t$ is pre-trained and acts as a generator ${G}$, which will conduct adversarial learning (minmax game) with the domain discriminator ${D}$ until both reach the optimal state.
\begin{align}
\small
    \min_{{F}_t}\max_{{D}}\mathcal{L}({D},{F}_t)&=\mathbb{E}_{\mathbf{x} \sim p_s(\mathbf{x})}[\log{D(F_s(\mathbf{x}))}] \notag \\
    &+ \mathbb{E}_{\mathbf{x} \sim p_t(\mathbf{x})}[\log{(1-D(F_t(\mathbf{x})))}]
    \label{loss}
\end{align}
Here $D$ is a simple binary perceptron used to distinguish source and target examples. The domain labels are set to 0 for the source samples and to 1 for the target samples. $D$ will score each input sample and try to minimize the error on both source and target tasks. For binary classification tasks, we choose the binary cross entropy function as the loss function, as shown in Eq. (\ref{loss}). On the other hand, the goal of domain adaptation is to minimize the difference in feature space between the source and target domains \cite{ben2007analysis}, so that the knowledge of the source domain can be successfully applied to the target domain. A good feature extractor $F_t$ or $G$ should be able to extract features that can confuse $D$, that is, maximizing the loss of $D$.

Once we have learned a good $F_s$, we fix and copy its parameters to $G$ as pre-trained \cite{tzeng2017adversarial}, and then iteratively train $G$ and $D$ until their losses no longer decrease. Following the same derivation as in GAN, when $D$ and $G$ are both optimal, the following relationship holds:
\begin{equation}
\small
    D^*(\mathbf z) = \frac{p_s(\mathbf z)}{p_s(\mathbf z) + p_t(\mathbf z)}
    \label{D}
\end{equation}
\begin{equation}
\small
    \mathcal L(F_t^*, D) = 2JS(p_s||p_t)-2\log 2 \ge -2\log 2
    \label{G}
\end{equation}

\textit{Proof.} We unify the feature spaces of the source and target domains transformed by the two feature extractors to get Eq. (\ref{max L}). We further set the partial derivative of $\mathcal{L}$ to $D$ as zero, and the optimal value of $D$ can be solved as Eq. (\ref{D}), where $p_s(\mathbf z)$ and $p_t(\mathbf z)$ are the marginal distributions in the new space. The purpose of training $G$ (domain adaptation) is to make the difference between the feature spaces as small as possible, so the ideal $G$ should make the $JS$ divergence close to 0, as shown in Eq. (\ref{G}).
\begin{align}
\small
    \mathcal{L} &= \int_{\mathbf{x}}p_s(\mathbf{x})\log(D(F_s(\mathbf{x}))) + p_t(\mathbf{x})\log(1-D(F_t(\mathbf{x}))) d\mathbf{x} \notag \\
    &= \int_{\mathbf{z}}p_s(\mathbf{z})\log(D(\mathbf z)) + p_t(\mathbf{z})\log(1-D(\mathbf z)) d\mathbf{z}
    \label{max L}
\end{align}

\begin{comment}
\begin{align}
    f(x) &= A\log x + B\log (1-x) \notag \\
    \frac{\mathrm{d} f}{\mathrm{d} x} &= A \frac{1}{\log} \frac{1}{x} - B \frac{1}{\log} \frac{1}{(1 - x)} \notag \\
    &= \frac{1}{\log} \frac{A-(A+B)x}{x(1-x)}
    \label{max D}
\end{align}
\begin{align}
    JS(p_s||p_t) &= \frac{1}{2}(KL(p_s||\frac{p_s+p_t}{2})+KL(p_t||\frac{p_s+p_t}{2})) \notag \\
     &= \frac{1}{2}(\log 2 + \int_{\mathbf{x}}p_s(\mathbf{x}) \log {\frac{p_s(\mathbf x)}{p_s(\mathbf x) + p_t(\mathbf x)}}) \notag \\
     &+ \frac{1}{2}(\log 2 + \int_{\mathbf{x}}p_t(\mathbf{x}) \log {\frac{p_t(\mathbf x)}{p_s(\mathbf x) + p_t(\mathbf x)}}) \notag \\
     &= \frac{1}{2}(2\log 2 + \mathcal{L}(D, F_t))
     \label{max G}
\end{align}
\end{comment}

Therefore, when the network training completed, for the classes shared by $\mathcal{D}_s$ and $\mathcal{D}_t$, their features should be able to confuse $D$. In this case, when $D$ affirmatively determines that a sample (class) comes from $\mathcal{D}_s$, this sample (class) should be unique to $\mathcal{D}_s$. In this way, we can infer the approximate classes of $\mathcal{D}_t$. $D$ will score all samples in $\mathcal{D}_s$, then calculate the average score for each class. The scores of unique classes in $\mathcal{D}_s$ are close to 0.

Since there may be source domains that have no intersection with $\mathcal{D}_t$, we remove the datasets whose classes are all unique, and re-execute the above steps to reduce negative transfer. We redefine those leftover datasets and scores as $\mathcal{D}_s=\{\mathcal{D}_{s_i}\}_{i=1}^{m}$ and $\{k_c\}_{c=1}^{n_y^s}$, where $m$ is the number of source datasets and $n_y^s$ is the label space size of $\mathcal{D}_s$.

Through this partial transfer learning, we can exclude the irrelevant source domains, roughly guess the probability $k_c$ of class $c$ within the target domain, and learn a feature extraction network $F_t \& F_s$ that extracts common features between domains. However, since some difficult samples may be difficult to be classified, so we provide an unknown class to accommodate these samples, then our next learning scenario can be modeled as an open set transfer learning, which can use $\{k_c\}_{c=1}^{n_y^s}$ as side information.
%{\color{red}[confusing $k_i$]}
%{\st{Unlike the standard open set transfer learning {\color{red}[`Given the comprehensive coverage of multiple source domains' can replace `unlike'? This is odd to say unlike, because we we work in a close set now?]}, there is no general class (outer class) composed of classes that are not of concern in the source or target domain.}}

\subsection{Open Set Transfer}
%To meet the third need (transfer knowledge from $\mathcal{D}_s$ to help model the tasks in $\mathcal{D}_t$),
In the common feature space learned by PDA, we assign class labels to samples of the target domain based on feature vector, respectively. These samples are either classified as one of the classes in $\mathcal{D}_s$ or as unknown. We classify by calculating the distance between the source domain sample and the center of each class in $\mathcal{D}_s$, similar to some metric-based meta learning work. The distance is calculated as:
\begin{equation}
\small
    d_{ci} = \frac{1}{k_c}||\mathbf{s}_c - \mathbf{x}_i||_2^2
    \label{dci}
\end{equation}
where $\mathbf{s}_c$ is the mean of all samples of class $c$ in the source domain with existence probability $k_c$. We assume that the sample is more likely to belong to the class with higher existence probability. We do not force each sample being assigned with a label; when the distance is larger than the max distance (Eq. \eqref{dmax}), we annotate the sample as unknown. We formulate the assignment problem as follows
\begin{equation}
\small
    \min_c d_{ci} \ \ \mathrm{ s.t. } \ d_{ci} \le \alpha d_{c}
    \label{assign}
\end{equation}
\begin{equation}
\small
    d_{c} = \max\{||\mathbf{s}_c - \mathbf{x}||_2^2\ |\ \mathbf{x} \in c\}
    \label{dmax}
\end{equation}
where hyper-parameter $\alpha$ is the relaxation coefficient. Through transfer learning, we obtain the approximate label set of the target domain and the rough classification results of samples therein. Those information will be used to guide the subsequent crowdsourcing process.

\subsection{Crowdsourcing}
\label{crowd}
Till now, we have transformed the open set crowdsourcing problem into a close set crowdsourcing problem with a label space size of $n_y^s$, where $n_y^s$ is the number of possible labels inferred by multiple source transfer learning. We divide the tasks into two pools: the task pool with machine annotations $\mathcal{D}_l$ and the task pool marked as unknown $\mathcal{D}_u$.
%{\color{red}We further add an option as ``other class'' to all tasks and allow crowd workers to type in the answer if they think there is no correct answer. [If so, how we really handle the open set?]}

The reason for this set up is closely related to our worker model. We believe that the characteristics of workers can be represented by two independent parameters, $(C, A)$. The binary parameter $C$ is conscientiousness obtained by analyzing the behavior characteristics of workers, representing the willingness of workers to do their best and tell the truth. The probability-like parameter $A$ represents the objective ability of workers, and it is obtained by calculating the accuracy of workers. If the worker’s subjective is insufficient, we will not accept their annotations regardless of the abilities, and we further distinguish honest workers based on their skill level.

In the crowdsourcing process, we keep three worker pools according to the worker model $(C, A)$: $\mathcal{W}_u$, $\mathcal{W}_r$, and $\mathcal{W}_e$. For example, the four types of crowd workers defined in \cite{kazai2011worker} (i.e. expert, normal, random, and spammer) can be grouped into random and spammer $\mathcal{W}_u$, normal $\mathcal{W}_r$, and expert $\mathcal{W}_e$. If the worker's answers are close to random guessing or always the same, the worker is considered as unreliable and put into $\mathcal{W}_u$. If a worker’s answer is of passable quality, we add the worker into $\mathcal{W}_r$. Further, if the worker’s answers are of high quality, and she/he has answered a sufficient number (\textit{e.g.} $\ge10$) of questions, then this worker is deemed as an expert and put into $\mathcal{W}_e$.
%or is willing to spend more time choosing ``other class'' and typing in the answers, then this worker is deemed as an expert and put into $\mathcal{W}_e$. This is in line with common sense: only workers who are responsible and skilled are willing to spend more time annotating a hard task while the reward remains unchanged \cite{kazai2011worker}.
%{\color{red}[One may concern why to choose 10?]}
%{\color{red}[any supportive references for this assumption?]}
%{\color{red}[Shall we make the crowdsourcing process too complicated? necessary? not shift our focus?]}

To make OSCrowd practical, we design our crowdsourcing process in an online scenario. In online scenarios, task allocation algorithms such as task-centering or worker-task selection will fail due to the uncontrollable availability of workers. Following the setting in \cite{ho2012online}, workers $\{w_i\}_{i=1}^{W}$ from a pool of size $W$ arrive one at a time and the task requester must assign a batch of tasks to the newly arrived worker. We don't know whether a worker will be available again, nor the order the workers arrive.

Our crowdsourcing process can be divided into two stages based on the level of understanding of workers. \\
\textbf{Stage 1 Explore:} When a worker arrives for the first time, we give priority to allocate tasks in $\mathcal{D}_l$, and ensure that each task can receive enough ($\ge 5$) annotations, so that we can infer the true label for the tasks and the worker capability. \\
\textbf{Stage 2 Collocate:} Case-1; to workers in $\mathcal{W}_u$ we still assign tasks in $\mathcal{D}_l$ that have received enough machine annotations. Case-2; to workers in $\mathcal{W}_r$ we assign tasks in $\mathcal{D}_l$ which do not have ascertained labels. Case-3; workers in $\mathcal{W}_e$ are assigned hard tasks in $\mathcal{D}_u$.
%{\color{red}[where the tasks really come from $\mathcal{D}_l$ or $\mathcal{D}_u$. AVOID vague statement for criticizing]}.

\begin{table*}[h!tbp]
\small
\centering
\begin{tabular}{c|c|c|cccc}
\hline
\multicolumn{3}{c|}{\textbf{Office31}}                         & \multicolumn{4}{c}{\textbf{OfficeHome}}                                                                                        \\ \hline
A (Target)   & D (Source) & W (Source) & \multicolumn{1}{c|}{A (Target)}   & \multicolumn{1}{c|}{C (Source)} & \multicolumn{1}{c|}{P (Source)} & R (Source)                  \\ \hline
\{0, 1, 2, 3, 4\} & \{0, 1\}, \{2, 5\} & \{3, 4, 6\}, \{7, 8\} & \multicolumn{1}{c|}{\{0, 1, 2, 3, 4\}} & \multicolumn{1}{c|}{\{0, 1, 5\}} & \multicolumn{1}{c|}{\{2, 3\}} & \{4, 6\}, \{7, 8\} \\ \hline
\end{tabular}
\vspace{-0.7em}
\caption{Composition of the crowdsourcing tasks. The source and target domains share classes $\{0,1,2,3,4\}$, but $\{5,6,7,8\}$ are unique to the source domain. Classes in `\{\ \}' constitute a dataset (domain). In Office31, we simulated four source domains from two distributions, and in OfficeHome four source domains from three distributions. \{7,8\} is completely unrelated source domain.}
\label{data}
\end{table*}

\begin{table}[h!tbp]
\small
\centering
\setlength{\tabcolsep}{1mm}{
    \begin{tabular}{c|c|ccccc|cccc}
\hline
                     & Score  & \textbf{0}    & \textbf{1}    & \textbf{2}    & \textbf{3}    & \textbf{4}    & 5    & 6    & 7    & 8    \\ \hline
\multirow{2}{*}{O31} & $k_c$-1 & \textbf{.687} & \textbf{.785} & \textbf{.849} & \textbf{.712} & \textbf{.823} & .411 & .523 & .237 & .341 \\
                     & $k_c$-2 & \textbf{.724} & \textbf{.765} & \textbf{.867} & \textbf{.723} & \textbf{.854} & .387 & .444 & -    & -    \\ \hline
\multirow{2}{*}{OH}  & $k_c$-1 & \textbf{.715} & \textbf{.732} & \textbf{.661} & \textbf{.849} & \textbf{.708} & .279 & .330 & .341 & .458 \\
                     & $k_c$-2 & \textbf{.696} & \textbf{.753} & \textbf{.744} & \textbf{.802} & \textbf{.776} & .264 & .375 & -    & -    \\ \hline
\end{tabular}
}
\vspace{-0.7em}
    \caption{The scores of each class in $\mathcal{D}_s$. We remove the irrelevant source domain (\{7,8\}) according to $k_c$ in round-1 ($k_c$-1), and run PDA again to get the $k_c$-2. It can be seen that scores of shared classes are significantly higher than those of unique classes.}
    \label{kc}
\end{table}

Since estimating task truth and worker ability usually requires abundant data to ensure quality, we do not aggregate the answers immediately if workers are available. Batch update is also feasible in the middle and late stage. To address the cold start problem, we first use the machine label as the ground truth, and update the worker ability ${A^w}$ of worker $w$ according to Eq. \eqref{acc} (number of correct annotations $N^w_{correct}$ divided by number of total annotations $N^w_{total}$ given by $w$). Then Eq. \eqref{trans} is used to soft $w$'s one-hot annotation $\mathbf{a}^w$ according to ${A^w}$. Next, we use Eq. \eqref{aggre} to aggregate annotations of task $t$ and select the label with the highest probability as the consensus label of $t$; $q^t$ is the number of total annotations received by task $t$. The EM algorithm is adopted to iteratively update task truth and worker model until convergence.
\begin{equation}
\small
    A^w = \frac{N_{correct}^w}{N_{total}^w} \times 100\%
    \label{acc}
\end{equation}
\begin{equation}
\small
    \tilde{\mathbf{a}}^w = Tr(\mathbf{a}^w | A^w) =
    \begin{cases}
        \frac{1-A^w}{n_y^s} & a^w_i=0  \\
        \ \ A^w & a^w_i=1 \\
    \end{cases}
    \label{trans}
\end{equation}
\begin{equation}
\small
    \hat{\mathbf{a}}^t = \frac{1}{q^t} {\sum}_{i=1}^{q^t} \tilde{\mathbf{a}}_i^t
    \label{aggre}
\end{equation}

In the \textit{Collocate} stage, we use Eq. (\ref{diver}) to measure the completion of a task. The greater the information entropy, the higher the uncertainty of the task is. So the greater $Com(\hat{\mathbf{a}}^t)$ in Eq. (\ref{diver}) is, the higher the completion of task $t$ (the denominator is the normalization of information entropy). When the completion of a task exceeds a certain threshold $\gamma$, or the number of annotations reaches the ceiling, we consider the task as completed.
\begin{equation}
\small
    Com(\hat{\mathbf{a}}^t) = 1 - \frac{H(\hat{\mathbf{a}}^t)}{\log n_y^s} = 1 + \frac{\sum_i^n p(\hat{a}_i^t) \log p(\hat{a}_i^t)}{\log n_y^s}
    \label{diver}
\end{equation}

Each time we update task label and worker model, we sort $\mathcal{D}_l$ according to the completion and adjust the worker pools. For workers identified in $\mathcal{W}_u$, their annotations do not have valid information, and we assign them completed tasks to confirm whether their evaluations need to be adjusted. If workers in $\mathcal{W}_r$ arrive, low completion tasks are assigned to them for further annotation. Finally, if an expert comes, we let this worker complete the most difficult tasks in $\mathcal{D}_u$. Since an expert’s annotation information is usually sufficient, each difficult task is annotated only once.

\section{Experiments}
\subsection{Settings}
Since OSCrowd is based on domain adaptation, we simulate our crowdsourcing tasks on two benchmark datasets in domain adaptation, \textit{Office31} (O31) \cite{saenko2010adapting} and \textit{OfficeHome} (OH) \cite{venkateswara2017deep}. Office31 is one of the commonly used datasets in domain adaptation. It has three sub-datasets, each containing the same 31 sub-classes collected from three different sources - Amazon (A), DSLR (D), and Webcam (W). As such, the sub-classes have different marginal distributions. Similarly, OfficeHome has four sub-datasets and 65 sub-classes extracted from four different styles - Art (A), Clipart (C), Product (P), and Real-world (R).

For Office31, we take five sub-classes in Amazon with the most samples as the annotation task. And then we select five identical sub-classes and four unrelated sub-classes from DSLR and Webcam, and combine them into four source domains. OfficeHome also uses a similar method to simulate multi-source domain open set crowdsourcing. See Table \ref{data} for the specific combinations. After PDA process, the scores of each class are shown in Table \ref{kc}, it can be seen that the target domain classes are well recognized out.

\begin{table}[ht]
\small
\centering
\begin{tabular}{c|c|c|ccc}
\hline
\textbf{Type} &$R$ &$A$    & \multicolumn{3}{c}{Ratio}                                    \\ \hline
expert        & 1          & {[}0.8,1.0{]} & \multicolumn{1}{c|}{10\%} & \multicolumn{1}{c|}{\textbf{20\%}} & 20\% \\
reliable      & 1          & {[}0.4,0.8{]} & \multicolumn{1}{c|}{70\%} & \multicolumn{1}{c|}{\textbf{60\%}} & 70\% \\
unreliable    & 0          & {[}0.1,0.4{]} & \multicolumn{1}{c|}{20\%} & \multicolumn{1}{c|}{\textbf{20\%}} & 10\% \\ \hline
\end{tabular}
\vspace{-0.7em}
\caption{Attributes and proportions of workers. $(R, A)$ are the attributes of the worker's model. The \textbf{boldfaced} ratio of workers is the baseline. The other two ratios are used to observe the model accuracy changes under different workers' quality.}
\label{worker}
\end{table}

With reference to \cite{kazai2011worker} and to our worker model, we simulate three different worker settings, and their models and percentage ratios are shown in Table \ref{worker}. We randomly take out one worker each time to simulate the arrivals of online workers.

\begin{table*}[h!tbp]
\small
\begin{center}
\setlength{\tabcolsep}{1.5mm}{
\begin{tabular}{c|c|c|c|cc|cc|c}
\hline
\textbf{Worker}  & \textbf{Data} & \textbf{WMV} & \textbf{GLAD} & \textbf{ALM} & \textbf{CMKT} & \textbf{DTA} & \textbf{RAOC} & \textbf{OSCrowd} \\ \hline
\multirow{2}{*}{Ratio-1} & O31  & 0.813$\pm$0.018 & 0.827$\pm$0.028 & 0.819$\pm$0.061 & 0.833$\pm$0.028 & 0.887$\pm$0.020 & 0.858$\pm$0.025 & 0.872$\pm$0.014         \\
                         & OH   & 0.817$\pm$0.013 & 0.823$\pm$0.020 & 0.815$\pm$0.057 & 0.824$\pm$0.025 & 0.878$\pm$0.019 & 0.843$\pm$0.020 & 0.868$\pm$0.016         \\ \hline
\multirow{2}{*}{\textbf{Ratio-2}} & O31  & 0.856$\pm$0.016 & 0.869$\pm$0.025 & 0.844$\pm$0.053 & 0.866$\pm$0.031 & 0.911$\pm$0.022 & 0.887$\pm$0.028 & 0.896$\pm$0.017         \\
                         & OH   & 0.840$\pm$0.011 & 0.855$\pm$0.019 & 0.838$\pm$0.048 & 0.851$\pm$0.021 & 0.906$\pm$0.016 & 0.877$\pm$0.018 & 0.884$\pm$0.013         \\ \hline
\multirow{2}{*}{Ratio-3} & O31  & 0.910$\pm$0.009 & 0.922$\pm$0.017 & 0.903$\pm$0.038 & 0.919$\pm$0.019 & 0.939$\pm$0.014 & 0.925$\pm$0.015 & 0.927$\pm$0.012         \\
                         & OH   & 0.902$\pm$0.008 & 0.917$\pm$0.019 & 0.888$\pm$0.035 & 0.905$\pm$0.020 & 0.925$\pm$0.017 & 0.911$\pm$0.017 & 0.918$\pm$0.011         \\ \hline
\end{tabular}}
\end{center}
\vspace{-0.7em}
\caption{Label accuracy and standard deviations of OSCrowd and compared methods on two datasets. Ratio-$i$ is related to the three worker settings in Table \ref{worker} and Ratio-2 is the baseline. Note that DTA assumes that the actual quality of the worker's annotations is known; as such its accuracy corresponds to the ideal scenario and upper bound.}
\label{acc-all}
\end{table*}

\begin{table}[h!tbp]
\small
\begin{center}
\setlength{\tabcolsep}{1.5mm}{
\begin{tabular}{c|c|ccccccc}
\hline
\multicolumn{2}{c|}{$\alpha$}     & 0.4   & 0.6   & 0.8   & 1.0   & \textbf{1.2}   & 1.4   & 1.6   \\ \hline
\multirow{3}{*}{O31} & p   & .841 & .829 & .815 & .800 & \textbf{.783} & .735 & .659 \\
                     & r   & .427 & .563 & .668 & .781 & \textbf{.856} & .882 & .902 \\ \cline{2-9}
                     & p*r & .359 & .467 & .544 & .625 & \textbf{.670} & .648 & .594 \\ \hline
\multirow{3}{*}{OH}  & p   & .778 & .762 & .739 & .708 & \textbf{.669} & .607 & .537 \\
                     & r   & .416 & .548 & .655 & .750 & \textbf{.828} & .864 & .884 \\ \cline{2-9}
                     & p*r & .324 & .418 & .484 & .531 & \textbf{.554} & .524 & .475 \\ \hline
\end{tabular}
}
\end{center}
\vspace{-0.7em}
\caption{The relationship between the precision (p) and recall (r) of the machine label with the relaxation coefficient $\alpha$ on two datasets.}
\label{alpha}
\end{table}

In order to verify the effectiveness of OSCrowd, we select four types of comparison methods: (i) crowdsourcing baseline, (ii) general crowdsourcing technology, (iii) crowdsourcing combining transfer learning and (iv) online crowdsourcing technology, with one or two representative methods for each type. Since the crowdsourcing methods below cannot handle open set annotations, we assume that the label space is known for these methods.

In crowdsourcing baseline, we assign five annotations to each task, and adopt Weighted Majority Voting and the EM algorithm to iteratively aggregate answers, which has been proved to be robust and effective in practical applications \cite{zheng2017truth}. We denote it as \textit{WMV}.

For the general crowdsourcing technology, \textit{GLAD} \cite{whitehill2009whose} uses a probabilistic approach to model the interaction between worker ability, task difficulty, and task ground truth, and jointly infer them to obtain final labels. We also fix the number of annotations to five for each task.

\textit{ALM} \cite{fang2014active} and \textit{CMKT} \cite{hancrowdsourcing} use transfer learning to obtain better task features. ALM transfers from a single source domain and CMKT uses multi-source transfer. ALM uses active learning to select the most informative tasks while CMKT uses task assignment based on worker's skill and task type. We use the default settings recommended by the original paper.

In an online scenario, \textit{DTA} \cite{ho2012online} assigns tasks with the goal of maximizing the total benefit that the requester obtains from the completed work. DTA first explores the worker's ability, then uses the primal-dual formulation to assign tasks. \textit{RAOC} \cite{welinder2010online} ranks workers' abilities and current tasks' confidence, then selects the task with the lowest confidence to annotate.

In OSCrowd, we set a task to obtain at most five annotations; we set the relaxation coefficient $\alpha=1.2$ and the completion threshold $\gamma=0.75$. Ten annotations are sufficient to infer a stable worker quality. Each dataset has more than $90 \times 5=450$ tasks, and we simulated 30 crowd workers. The related experimental results are shown in Table \ref{acc-all}.
%{\color{red}[It is better to specify the parameters of compared methods. We mentioned weighs source domains, but not well explain it?]}

\subsection{Analysis of Results}
By comparing the above methods, we have the following important observations: \\
(i) \textbf{OSCrowd vs. Online:} DTA achieves the best results among all methods and OSCrowd achieves sub-optimal results. However, this is because DTA assumes that the quality of workers is known without inference. As such, DTA is more like an optimal planning algorithm rather than a crowdsourcing algorithm. Therefore, the quality achieved by DTA gives an upper bound for the other methods. Although OSCrowd and RAOC have similar problem models, with the help of machine labels given by transfer learning, OSCrowd can better deal with the initial unstable stage of the model, and its overall quality is higher than that of RAOC. \\
(ii) \textbf{OSCrowd vs. Transfer:} Both ALM and CMKT use transfer learning to assist crowdsourcing modeling. CMKT transfers knowledge from multi-source domains, which reduces the risk of negative transfer from a single source domain, thus achieving higher accuracy than ALM. OSCrowd further calculates the correlation of multiple source domains, excludes unrelated source domains to reduce the risk of negative transfer, and thus achieves a better performance. When using transfer learning, avoiding negative transfer is an important way to improve model quality. \\
(iii) \textbf{OSCrowd vs. GLAD \& WMV:} Both GLAD and WMV use the `accuracy' to model crowd workers. GLAD further considers the correction effect of task difficulty on worker ability, thus achieving higher accuracy than WMV. However, they do not selectively assign tasks to workers. OSCrowd tries to grade workers and tasks, then assigns simple tasks to ordinary workers and leaves difficult tasks to experts, discarding the annotations from unreliable workers, so it achieves better quality than both GLAD and WMV. \\
(iv) \textbf{Impact of different worker quality:} Observing Table \ref{acc-all} Ratio-3, when the proportion of low-quality workers decreases, almost all methods achieve better quality, and the performance gap between them is eliminated. This shows that all seven methods are effective crowdsourcing algorithms, and that the most important factor in determining the quality of real crowdsourcing is still the crowd workers. Conversely, when the quality of workers decreases (Table \ref{acc-all} Ratio-1), methods containing tasks assignment design are less affected (all methods except WMV and GLAD), because they can make better use of high-quality workers and reduce noise impact caused by low-quality workers.

\subsection{Ablation Study}
We studied the impact of the relaxation coefficient $\alpha$ in Eq. (\ref{assign}) and the completion threshold $\gamma$ related to Eq. (\ref{diver}). \cite{sheng2008get} has studied the effect of repeated annotations. The investigation of how many annotations lead to a stable worker model in section \ref{crowd} is a purely statistical question.

In general, a smaller $\alpha$ will bring higher machine label accuracy, but at the same time more samples will be labeled as unknown (difficult task $D_u$). We borrow precision and recall to express this relationship, and adopt their product to consider both factors, the results are shown in Table \ref{alpha}. It can be seen that for Office31, the optimal value is between 1.2 and 1.4, and for OfficeHome, it's between 1.0 and 1.2, therefore we choose $\alpha=1.2$.

As for $\gamma$, there is a specific relationship between completion and the expected label quality, bigger $\gamma$ leads to better quality and more budget. We expect the label quality to be about 0.9 under five classes. The two extreme cases of task annotation are $(0.9, 0.1, 0, 0, 0)$ and $(0.9, 0.025, 0.25, 0.025, 0.025)$, then the corresponding completion can be calculated out as 0.798 and 0.712 respectively, so we eclectically set $\gamma=0.75$.

\section{Conclusion}
We propose and define open set crowdsourcing annotation, and analyze its difficulties and the shortcomings of current methods. We propose to use two-stage transfer learning to help solve this problem: OSCrowd first uses multi-source adversarial partial domain adaptation to infer the possible label space of crowdsourcing tasks, and excludes irrelevant source domains, and then labels crowdsourcing tasks. The label space information and machine labels provided by transfer learning are used to assist the crowdsourcing design. For crowdsourcing, OSCrowd scores crowdsourcing workers and assigns tasks based on their abilities and commitment. Simulation experiments have confirmed that our method effectively solves the open set crowdsourcing annotation problem.

\bibliographystyle{named}
\bibliography{Crowd_Bib}

\begin{thebibliography}{}

\bibitem[\protect\citeauthoryear{Ben-David \bgroup \em et al.\egroup
  }{2007}]{ben2007analysis}
Shai Ben-David, John Blitzer, Koby Crammer, and Fernando Pereira.
\newblock Analysis of representations for domain adaptation.
\newblock In {\em NeurIPS}, pages 137--144, 2007.

\bibitem[\protect\citeauthoryear{Blitzer \bgroup \em et al.\egroup
  }{2006}]{blitzer2006domain}
John Blitzer, Ryan McDonald, and Fernando Pereira.
\newblock Domain adaptation with structural correspondence learning.
\newblock In {\em EMNLP}, pages 120--128, 2006.

\bibitem[\protect\citeauthoryear{Cao \bgroup \em et al.\egroup
  }{2019}]{cao2019learning}
Zhangjie Cao, Kaichao You, Mingsheng Long, Jianmin Wang, and Qiang Yang.
\newblock Learning to transfer examples for partial domain adaptation.
\newblock In {\em CVPR}, pages 2985--2994, 2019.

\bibitem[\protect\citeauthoryear{Fang \bgroup \em et al.\egroup
  }{2014}]{fang2014active}
Meng Fang, Jie Yin, and Dacheng Tao.
\newblock Active learning for crowdsourcing using knowledge transfer.
\newblock In {\em AAAI}, pages 1809--1815, 2014.

\bibitem[\protect\citeauthoryear{Ganin and
  Lempitsky}{2015}]{ganin2015unsupervised}
Yaroslav Ganin and Victor Lempitsky.
\newblock Unsupervised domain adaptation by backpropagation.
\newblock In {\em ICML}, pages 1180--1189, 2015.

\bibitem[\protect\citeauthoryear{Ganin \bgroup \em et al.\egroup
  }{2016}]{ganin2016domain}
Yaroslav Ganin, Evgeniya Ustinova, Hana Ajakan, Pascal Germain, Hugo
  Larochelle, Fran{\c{c}}ois Laviolette, Mario Marchand, and Victor Lempitsky.
\newblock Domain-adversarial training of neural networks.
\newblock {\em JMLR}, 17(1):2096--2030, 2016.

\bibitem[\protect\citeauthoryear{Goodfellow \bgroup \em et al.\egroup
  }{2014}]{goodfellow2014generative}
Ian Goodfellow, Jean Pouget-Abadie, Mehdi Mirza, Bing Xu, David Warde-Farley,
  Sherjil Ozair, Aaron Courville, and Yoshua Bengio.
\newblock Generative adversarial nets.
\newblock In {\em NeurIPS}, pages 2672--2680, 2014.

\bibitem[\protect\citeauthoryear{Haas \bgroup \em et al.\egroup
  }{2015}]{haas2015clamshell}
Daniel Haas, Jiannan Wang, Eugene Wu, and Michael~J Franklin.
\newblock Clamshell: Speeding up crowds for low-latency data labeling.
\newblock {\em VLDB Endowment}, 9(4):372--383, 2015.

\bibitem[\protect\citeauthoryear{Han \bgroup \em et al.\egroup
  }{2020}]{hancrowdsourcing}
Guangyang Han, Jinzheng Tu, Guoxian Yu, Jun Wang, and Carlotta Domeniconi.
\newblock Crowdsourcing with multiple-source knowledge transfer.
\newblock {\em IJCAI}, pages 2908--2914, 2020.

\bibitem[\protect\citeauthoryear{Ho and Vaughan}{2012}]{ho2012online}
Chien-Ju Ho and Jennifer Vaughan.
\newblock Online task assignment in crowdsourcing markets.
\newblock In {\em AAAI}, pages 45--51, 2012.

\bibitem[\protect\citeauthoryear{Ho \bgroup \em et al.\egroup
  }{2013}]{ho2013adaptive}
Chien-Ju Ho, Shahin Jabbari, and Jennifer~Wortman Vaughan.
\newblock Adaptive task assignment for crowdsourced classification.
\newblock In {\em ICML}, pages 534--542, 2013.

\bibitem[\protect\citeauthoryear{Hoffman \bgroup \em et al.\egroup
  }{2018}]{hoffman2018cycada}
Judy Hoffman, Eric Tzeng, Taesung Park, Jun-Yan Zhu, Phillip Isola, Kate
  Saenko, Alexei Efros, and Trevor Darrell.
\newblock Cycada: Cycle-consistent adversarial domain adaptation.
\newblock In {\em ICML}, pages 1989--1998, 2018.

\bibitem[\protect\citeauthoryear{Kazai \bgroup \em et al.\egroup
  }{2011}]{kazai2011worker}
Gabriella Kazai, Jaap Kamps, and Natasa Milic-Frayling.
\newblock Worker types and personality traits in crowdsourcing relevance
  labels.
\newblock In {\em CIKM}, pages 1941--1944, 2011.

\bibitem[\protect\citeauthoryear{Li \bgroup \em et al.\egroup
  }{2016}]{li2016crowdsourcing}
Qi~Li, Fenglong Ma, Jing Gao, Lu~Su, and Christopher~J Quinn.
\newblock Crowdsourcing high quality labels with a tight budget.
\newblock In {\em WSDM}, pages 237--246, 2016.

\bibitem[\protect\citeauthoryear{Li \bgroup \em et al.\egroup
  }{2017}]{li2017crowdsourced}
Guoliang Li, Yudian Zheng, Ju~Fan, Jiannan Wang, and Reynold Cheng.
\newblock Crowdsourced data management: Overview and challenges.
\newblock In {\em SIGMOD}, pages 1711--1716, 2017.

\bibitem[\protect\citeauthoryear{Long \bgroup \em et al.\egroup
  }{2013}]{long2013transfer}
Mingsheng Long, Jianmin Wang, Guiguang Ding, Jiaguang Sun, and Philip~S Yu.
\newblock Transfer feature learning with joint distribution adaptation.
\newblock In {\em ICCV}, pages 2200--2207, 2013.

\bibitem[\protect\citeauthoryear{Panareda~Busto and
  Gall}{2017}]{panareda2017open}
Pau Panareda~Busto and Juergen Gall.
\newblock Open set domain adaptation.
\newblock In {\em ICCV}, pages 754--763, 2017.

\bibitem[\protect\citeauthoryear{Saenko \bgroup \em et al.\egroup
  }{2010}]{saenko2010adapting}
Kate Saenko, Brian Kulis, Mario Fritz, and Trevor Darrell.
\newblock Adapting visual category models to new domains.
\newblock In {\em ECCV}, pages 213--226. Springer, 2010.

\bibitem[\protect\citeauthoryear{Sheng \bgroup \em et al.\egroup
  }{2008}]{sheng2008get}
Victor~S Sheng, Foster Provost, and Panagiotis~G Ipeirotis.
\newblock Get another label? improving data quality and data mining using
  multiple, noisy labelers.
\newblock In {\em KDD}, pages 614--622, 2008.

\bibitem[\protect\citeauthoryear{Tu \bgroup \em et al.\egroup
  }{2020}]{Tu2020CrowdWT}
Jinzheng Tu, Guoxian Yu, Jun Wang, Carlotta Domeniconi, Maozu Guo, and
  Xiangliang Zhang.
\newblock Crowdwt: Crowdsourcing via joint modeling of workers and tasks.
\newblock {\em ACM TKDD}, 15(1):1--24, 2020.

\bibitem[\protect\citeauthoryear{Tzeng \bgroup \em et al.\egroup
  }{2014}]{tzeng2014deep}
Eric Tzeng, Judy Hoffman, Ning Zhang, Kate Saenko, and Trevor Darrell.
\newblock Deep domain confusion: Maximizing for domain invariance.
\newblock {\em arXiv preprint arXiv:1412.3474}, 2014.

\bibitem[\protect\citeauthoryear{Tzeng \bgroup \em et al.\egroup
  }{2017}]{tzeng2017adversarial}
Eric Tzeng, Judy Hoffman, Kate Saenko, and Trevor Darrell.
\newblock Adversarial discriminative domain adaptation.
\newblock In {\em CVPR}, pages 7167--7176, 2017.

\bibitem[\protect\citeauthoryear{Venkateswara \bgroup \em et al.\egroup
  }{2017}]{venkateswara2017deep}
Hemanth Venkateswara, Jose Eusebio, Shayok Chakraborty, and Sethuraman
  Panchanathan.
\newblock Deep hashing network for unsupervised domain adaptation.
\newblock In {\em CVPR}, pages 5018--5027, 2017.

\bibitem[\protect\citeauthoryear{Welinder and
  Perona}{2010}]{welinder2010online}
Peter Welinder and Pietro Perona.
\newblock Online crowdsourcing: rating annotators and obtaining cost-effective
  labels.
\newblock In {\em CVPR}, pages 25--32, 2010.

\bibitem[\protect\citeauthoryear{Whitehill \bgroup \em et al.\egroup
  }{2009}]{whitehill2009whose}
Jacob Whitehill, Ting-fan Wu, Jacob Bergsma, Javier Movellan, and Paul Ruvolo.
\newblock Whose vote should count more: Optimal integration of labels from
  labelers of unknown expertise.
\newblock In {\em NeurIPS}, pages 2035--2043, 2009.

\bibitem[\protect\citeauthoryear{Zhang \bgroup \em et al.\egroup
  }{2018}]{zhang2018importance}
Jing Zhang, Zewei Ding, Wanqing Li, and Philip Ogunbona.
\newblock Importance weighted adversarial nets for partial domain adaptation.
\newblock In {\em CVPR}, pages 8156--8164, 2018.

\bibitem[\protect\citeauthoryear{Zheng \bgroup \em et al.\egroup
  }{2017}]{zheng2017truth}
Yudian Zheng, Guoliang Li, Yuanbing Li, Caihua Shan, and Reynold Cheng.
\newblock Truth inference in crowdsourcing: Is the problem solved?
\newblock {\em VLDB Endowment}, 10(5):541--552, 2017.

\bibitem[\protect\citeauthoryear{Zhuang \bgroup \em et al.\egroup
  }{2020}]{zhuang2020comprehensive}
Fuzhen Zhuang, Zhiyuan Qi, Keyu Duan, Dongbo Xi, Yongchun Zhu, Hengshu Zhu, Hui
  Xiong, and Qing He.
\newblock A comprehensive survey on transfer learning.
\newblock {\em Proceedings of the IEEE}, 99(1):43--76, 2020.

\end{thebibliography}

\end{document}